\documentclass[conference]{IEEEtran}
\ifCLASSINFOpdf

\else

   \usepackage{graphicx}
   \graphicspath{{../eps/}}
   \DeclareGraphicsExtensions{.eps}
\fi

\usepackage{amsmath}
\usepackage{amsfonts,amssymb}
\usepackage{amssymb}
\usepackage{wrapfig}
\usepackage{psfrag}
\usepackage{epstopdf}
\usepackage{cite}
\usepackage{graphicx}
\usepackage{subfigure}
\usepackage{threeparttable}
\usepackage{cases}
\usepackage{subeqnarray}
\usepackage{color}
\usepackage{underscore}
\usepackage{verbatim}
\usepackage{bm}

\begin{document}

\title{Enabling Panoramic Full-Angle Reflection via Aerial Intelligent Reflecting Surface}

\author{\IEEEauthorblockN{$\text{Haiquan~Lu}^*$, $\text{Yong~Zeng}^*$, $\text{Shi~Jin}^*$, and $\text{Rui~Zhang}^\dagger$ }
\IEEEauthorblockA{*National Mobile Communications Research Laboratory, Southeast University, Nanjing 210096, China\\
$\dagger$Department of Electrical and Computer Engineering, National University of Singapore, Singapore 117583\\
Email: haiq_lu@163.com, yong_zeng@seu.edu.cn, jinshi@seu.edu.cn, elezhang@nus.edu.sg}
}
\maketitle

\begin{abstract}
  This paper proposes a new three dimensional (3D) networking architecture enabled by aerial intelligent reflecting surface (AIRS) to achieve panoramic signal reflection from the sky. Compared to the conventional terrestrial IRS, AIRS not only enjoys higher deployment flexibility, but also is able to achieve 360$^\circ$ panoramic full-angle reflection and requires fewer reflections in general due to its higher likelihood of having line of sight (LoS) links with the ground nodes. We focus on the problem to maximize the worst-case signal-to-noise ratio (SNR) in a given coverage area by jointly optimizing the transmit beamforming, AIRS placement and phase shifts. The formulated problem is non-convex and the optimization variables are coupled with each other in an intricate manner. To tackle this problem, we first consider the special case of single-location SNR maximization to gain useful insights, for which the optimal solution is obtained in closed-form. Then for the general case of area coverage, an efficient suboptimal solution is proposed by exploiting the similarity between phase shifts optimization for IRS and analog beamforming for the conventional phase array. Numerical results show that the proposed design can achieve significant performance gain than heuristic AIRS deployment schemes.
\end{abstract}

\IEEEpeerreviewmaketitle
\section{Introduction}
Wireless communication aided by intelligent reflecting surface (IRS) has been proposed as a promising technology to realize energy-efficient and cost-effective transmissions in the future \cite{Intelligent walls as autonomous parts of smart indoor environments, Wireless communications with programmable metasurface Transceiver design and experimental results, Beamforming optimization for wireless network aided by intelligent reflecting surface with discrete phase shifts,Large intelligent surface-assisted wireless communication exploiting statistical CSI, Wireless communications with reconfigurable intelligent surface Path loss modeling and experimental measurement, Increasing indoor spectrum sharing capacity using smart reflect-array, Intelligent reflecting surface meets OFDM protocol design and rate maximization,Wireless communications through reconfigurable intelligent surfaces,Achievable rate maximization by passive intelligent mirrors, Towards smart and reconfigurable environment: Intelligent reflecting surface aided wireless network, Intelligent reflecting surface enhanced wireless network via joint active and passive beamforming,Intelligent reflecting surface vs. decode-and-forward How large surfaces are needed to beat relaying}. IRS is a man-made reconfigurable metasurface composed of a large number of regularly arranged passive reflecting elements and a smart controller \cite{Towards smart and reconfigurable environment: Intelligent reflecting surface aided wireless network}. Through modifying the amplitude and/or phase shift of the radio signal impinging upon its reflecting elements, IRS is able to achieve highly accurate radio wave manipulation in desired manners, which thus offers a new paradigm of wireless communication system design via controlling the radio propagation environment for various purposes, such as  signal enhancement, interference suppression and transmission security\cite{Towards smart and reconfigurable environment: Intelligent reflecting surface aided wireless network}. Thanks to the passive array architecture, IRS-aided wireless communication is able to reap the benefits of large antenna arrays with low power consumption and hardware cost. Furthermore, different from the conventional relays, the radio signal reflected by IRS is free from self-interference and noise in an inherently full-duplex transmission manner \cite{Towards smart and reconfigurable environment: Intelligent reflecting surface aided wireless network}.

\begin{figure}[!t]
    \setlength{\abovecaptionskip}{-0.05cm}
  \setlength{\belowcaptionskip}{-0.3cm}
  \centering
  \subfigure[Terrestrial IRS]{
    \includegraphics[width=1.2in,height=0.95in]{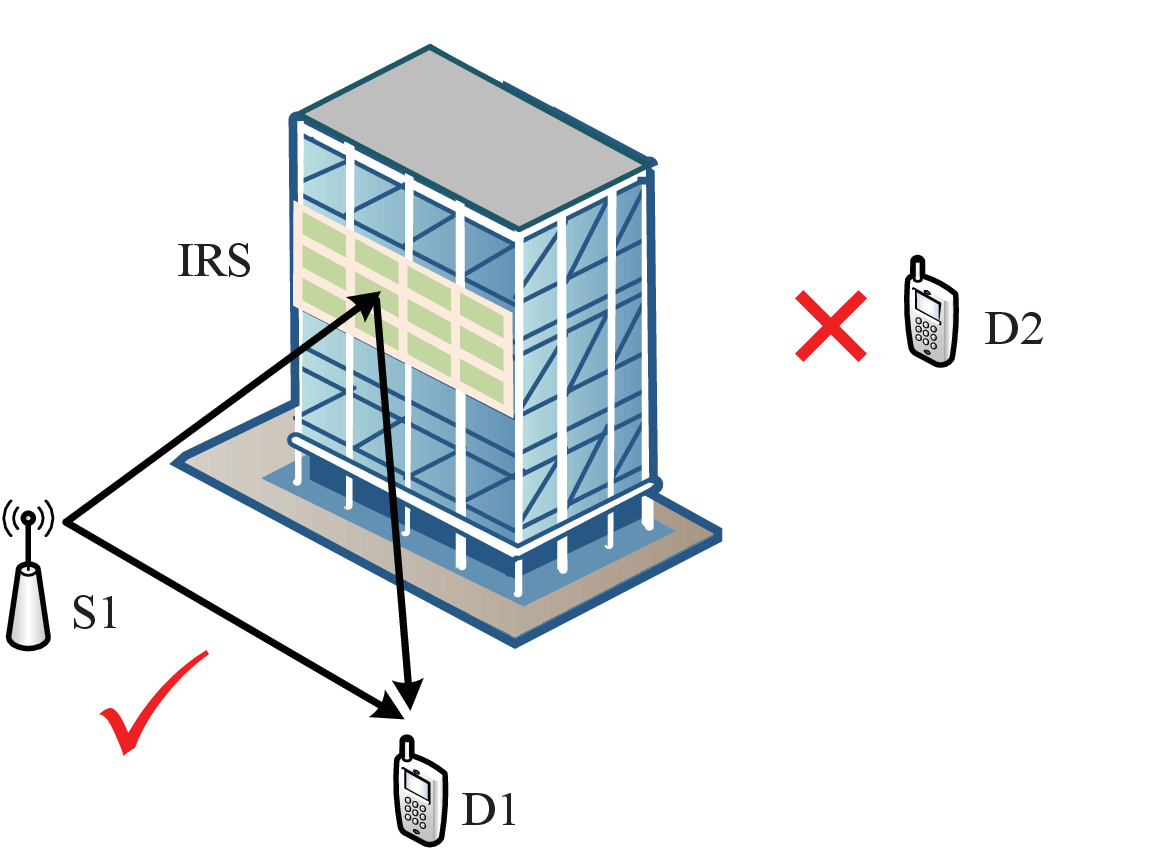}
  }
  \hspace{10mm}
  \subfigure[AIRS]{
    \includegraphics[width=1.2in,height=1.3in]{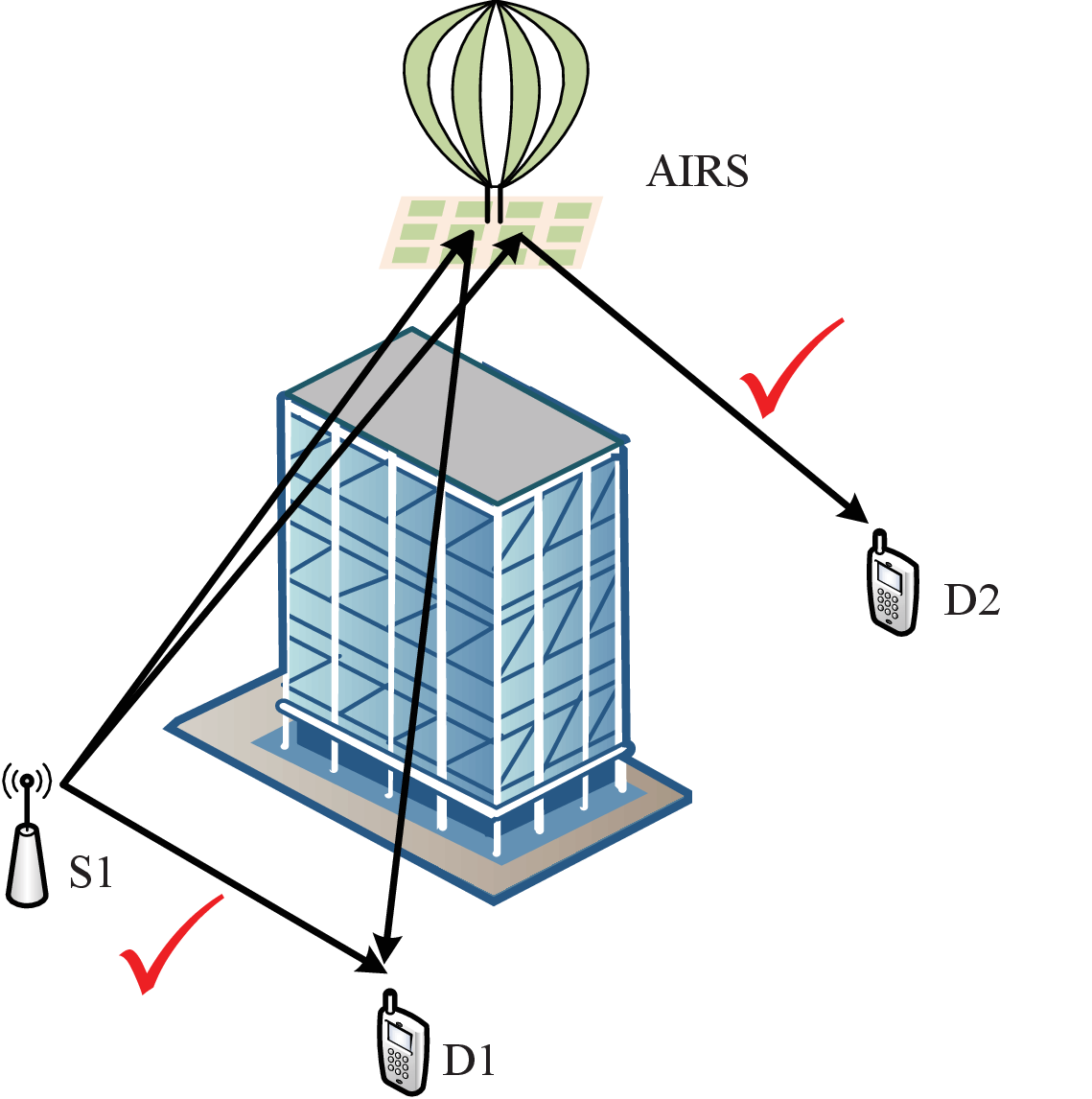}
  }
  \caption{180$^\circ$ half-space reflection by terrestrial IRS versus 360$^\circ$ panoramic full-angle reflection by AIRS.}
  \label{half and full of the space}
    \vspace{-0.6cm}
  \end{figure}

\begin{figure}[!t]
    \setlength{\abovecaptionskip}{-0.05cm}
  \setlength{\belowcaptionskip}{-0.3cm}
  \centering
  \subfigure[Terrestrial IRS]{
    \includegraphics[width=1.2in,height=1.1in]{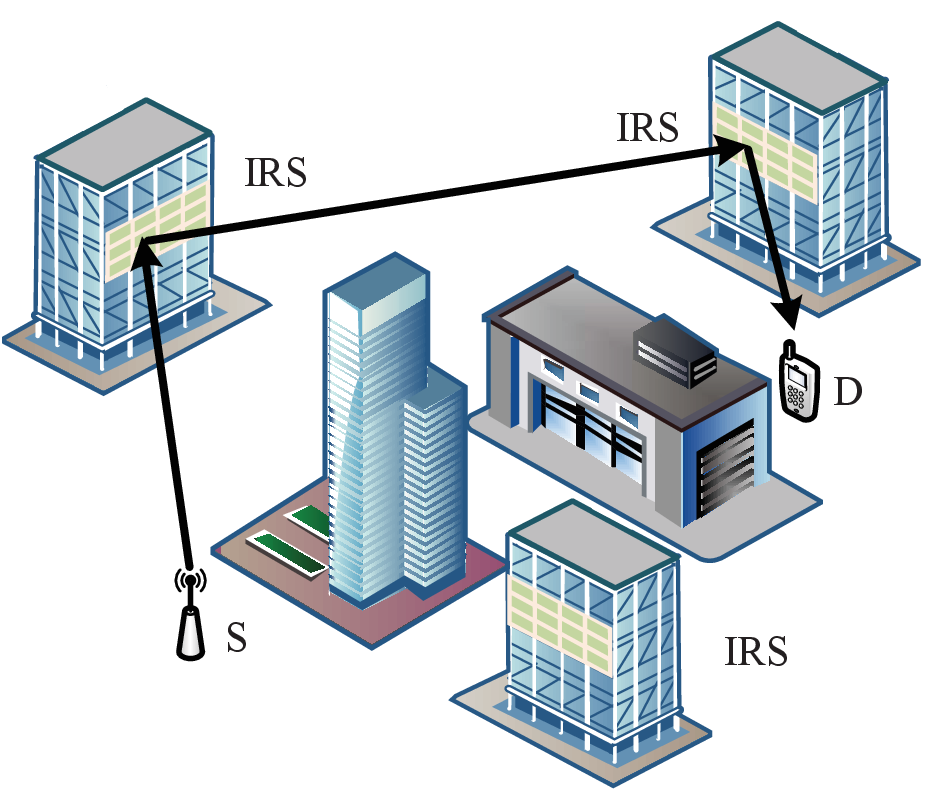}
  }
  \hspace{10mm}
  \subfigure[AIRS]{
    \includegraphics[width=1.2in,height=1.35in]{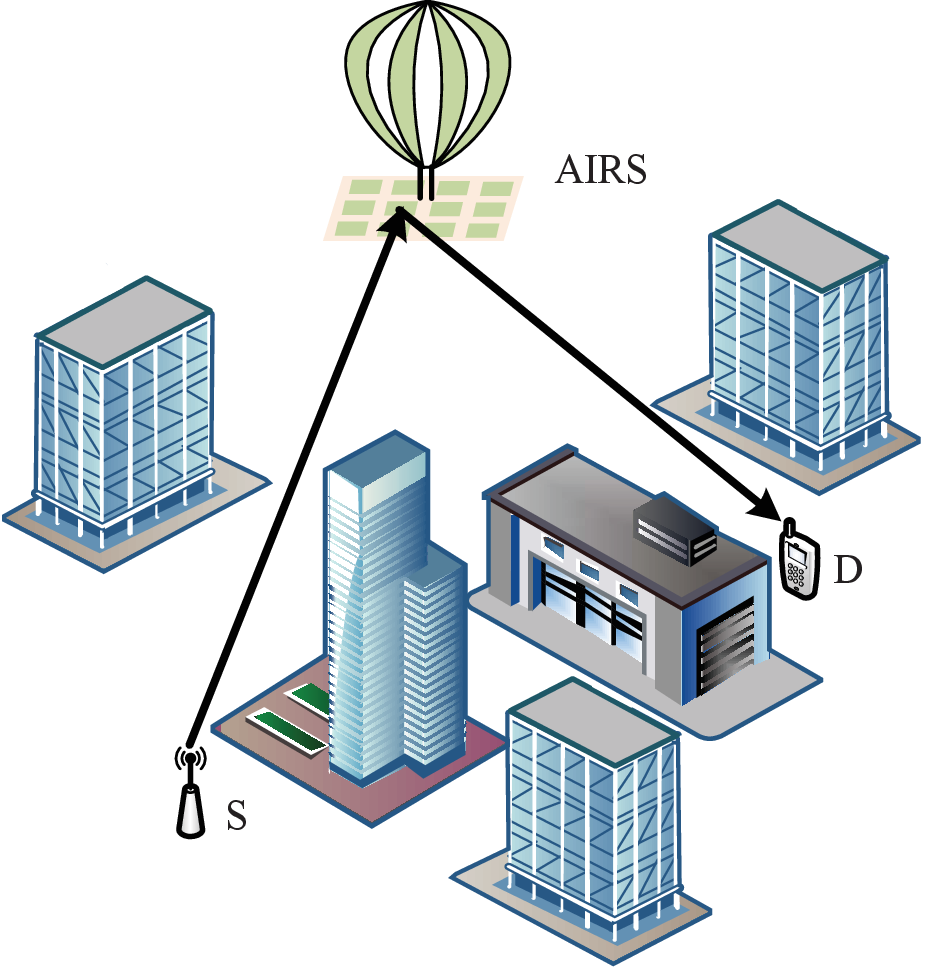}
  }
  \caption{AIRS reduces the number of reflections than terrestrial IRS.}
  \label{Different reflection times between IRS deployed on the facade of building and aerial platforms}
    \vspace{-0.6cm}
  \end{figure}

Most existing research mainly focuses on terrestrial IRS deployed on facades of buildings or indoor walls, which, however,  poses fundamental performance limitations for several reasons. First, from the deployment perspective, finding the appropriate place for IRS installation is usually a difficult task in practice. The installation process may also involve other issues, e.g., site rent, impact of urban landscape and the willingness of owners to install large IRS on their properties. Second, from the performance perspective, IRS deployed on the walls or facades of buildings can at most serve terminals located in half of the space, i.e., both the source and destination nodes must lie on the same side of the IRS, as illustrated in Fig.~\ref{half and full of the space}(a). Third, as shown in Fig.~\ref{Different reflection times between IRS deployed on the facade of building and aerial platforms}(a), in complex environment like urban areas, the radio signal originated from a source node has to be reflected many times before reaching the desired destination node, even with the presence of sufficient number of IRSs. This thus leads to significant signal attenuation since each reflection, even by IRS, would cause signal scattering to undesired directions. 
  
To address the above issues, we propose in this paper a novel three dimensional (3D) networking architecture enabled by aerial IRS (AIRS), for which IRS is mounted on aerial platforms like balloons, unmanned aerial vehicles (UAVs), so as to enable intelligent reflection from the sky. Compared to the conventional terrestrial IRS, AIRS has several appealing advantages. First, with elevated position, AIRS is able to establish line-of-sight (LoS) links with the ground nodes with high probability~\cite{Accessing from the sky A tutorial on UAV communications for 5G and beyond}, which leads to stronger channel as compared to the terrestrial IRS. At the same time, the placement or trajectory of aerial platforms can be more flexibly optimized to further improve the communication performance, thereby offering a new degree of freedom (DoF) for performance enhancement via 3D network design. Second, AIRS is able to achieve 360$^\circ$ panoramic full-angle reflection, i.e., one AIRS can in principle manipulate signals between any pair of nodes located on the ground, as illustrated in Fig.~\ref{half and full of the space}(b). This is in a sharp contrast to the conventional terrestrial IRS that usually can only serve nodes in half of the space. Last but not least, in contrast to the terrestrial IRS, AIRS is usually able to achieve desired signal manipulation by one reflection only, even in complex urban environment (see Fig.~\ref{Different reflection times between IRS deployed on the facade of building and aerial platforms}(b)), thanks to its high likelihood of having LoS links with the ground nodes. This thus greatly reduces the signal power loss due to multiple reflections in the case of terrestrial IRS. 

In this paper, we consider a basic setup of an AIRS-assisted communication system, where an AIRS is deployed to enhance the signal coverage of a given target area, say, a hot spot in the cellular network or a remote area without cellular coverage. Our objective is to maximize the minimum achievable signal-to-noise ratio (SNR) for the target area by jointly optimizing the transmit beamforming of the source node, the placement and phase shifts of the AIRS. The formulated problem is non-convex and difficult to be optimally solved in general. To gain useful insights at first, we consider the special case of single-location SNR maximization problem, for which the optimal AIRS placement and phase shifts are derived in closed-form. In particular, the optimal location of the AIRS is shown to only depend on the ratio between the AIRS height and the source-destination distance. For the general case of area coverage, we propose an efficient design by exploiting the similarity between phase shifts optimization for IRS and analog beamforming for the conventional phase array. Numerical results are presented which show the significant performance gain of the proposed design as compared to heuristic AIRS deployment schemes.

% >>>>>>>>>>>>>SECTIONS II -  here >>>>>>>>>>>>
\section{System Model And Problem Formulation}
 \begin{figure}[!t]
     \setlength{\abovecaptionskip}{-0.2cm}
  \setlength{\belowcaptionskip}{-0.4cm}
  \centering
  \centerline{\includegraphics[width=3.0in,height=1.80in]{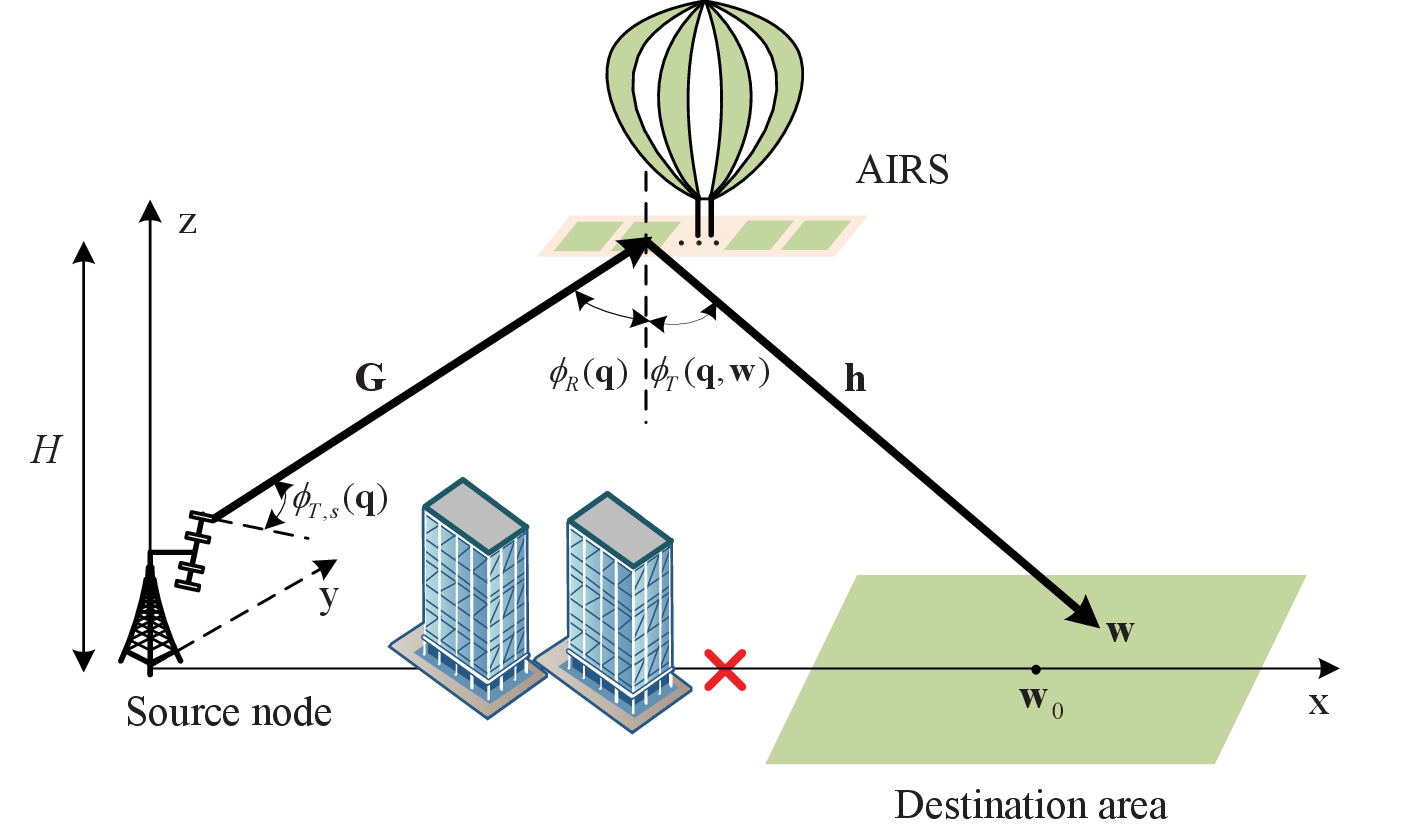}}
  \caption{AIRS-assisted wireless communication system.}
  \label{system model}
      \vspace{-0.6cm}
  \end{figure}
  As illustrated in Fig.~\ref{system model}, we consider an AIRS-assisted wireless communication system, where an AIRS is deployed to assist the source node (say a ground base station, access point or a user terminal) to enhance its communication performance within a given area $\mathcal A$ (assumed to be rectangular for the purpose of exposition). We assume that the direct communication link from the source node to the target area is negligible due to severe blockage. The source node is equipped with $M$ transmit antennas, where the adjacent antenna elements are separated by $d_0$. The AIRS comprises of a uniform linear array (ULA) with $N$ passive reflecting elements, separated by the distance $d < \lambda $, where $\lambda$ is the carrier wavelength. Without loss of generality, we assume that the source node is located at the origin in a Cartesian coordinate system and the center of the coverage area is on the x-axis, which is denoted by ${{\bf{w}}_0} = {\left[ {{x_0},0} \right]^T}$. Therefore, any location in the rectangular area $\mathcal A$ can be specified as ${\bf{w}} = {\left[ {{x_a},{y_a}} \right]^T}, {x_a} \in \left[ {{x_0} - \frac{{{D_x}}}{2},{x_0} + \frac{{{D_x}}}{2}} \right], {y_a} \in \left[ { - \frac{{{D_y}}}{2},\frac{{{D_y}}}{2}} \right]$, with $D_x$  and $D_y$ denoting the length and width of the rectangular area, respectively. For convenience, we assume $D_x\geq D_y$.

  The AIRS is assumed to be placed at an altitude $H$. In addition, consider the first reflection element of the AIRS as the reference point, whose horizontal coordinate is denoted by ${{\bf{q}}} = {\left[ {{x},{y}} \right]^T}$.  Therefore, the distance from the source node to the AIRS, and that from the AIRS to any location in $\mathcal A$ can be expressed as {\small{${d_{\bf{G}}} = \sqrt {{H^2} + {{\left\| {{{\bf{q}}}} \right\|}^2}} $}} and {\small{${d_{\bf{h}}} = \sqrt {{H^2} + {{\left\| {{{\bf{q}}} - {\bf{w}}} \right\|}^2}} $}}, respectively.
  
  In practice, the communication links between the aerial platform and ground nodes are LoS with high probability. Thus, for simplicity, we assume that the channel power gains follow the free-space path loss model, and the channel power gain from the source node to the AIRS can be expressed as 
  \begin{equation}
{\beta _{\bf{G}}}\left( {\bf{q}} \right) = \frac{{{\beta _0}}}{{{H^2} + {{\left\| {\bf{q}} \right\|}^2}}},
  \setlength\abovedisplayskip{1pt}
 \setlength\belowdisplayskip{1pt}
 \end{equation}
 where ${\beta _{\rm{0}}}$ represents the channel power at the reference distance ${d_0} = 1$~m.  Similarly, the channel power gain from the AIRS to a location $\mathbf w\in \mathcal A$ can be expressed as 
 \begin{equation}
  \setlength\abovedisplayskip{1pt}
 \setlength\belowdisplayskip{1pt}
{\beta _{\bf{h}}}\left( {{\bf{q}},{\bf{w}}} \right) = \frac{{{\beta _0}}}{{{H^2} + {{\left\| {{\bf{q}} - {\bf{w}}} \right\|}^2}}}.
 \end{equation}

 Let ${{\phi _{T,s}}\left( {\bf{q}} \right)}$ and ${{\phi _R}\left( {\bf{q}} \right)}$ be the angle of departure (AoD) and angle of arrival (AoA) of the signal from the source node to the AIRS, respectively. Then the channel matrix from the source node to the AIRS, denoted as ${\bf{G}}\left( {\bf{q}} \right) \in {{\mathbb{C}}^{N \times M}}$, can be expressed as 
   \begin{equation}
    \setlength\abovedisplayskip{1pt}
 \setlength\belowdisplayskip{1pt}
{\bf{G}}\left( {\bf{q}} \right) = \sqrt {{\beta _{\bf{G}}}\left( {\bf{q}} \right)} {e^{ - j\frac{{2\pi {d_{\bf{G}}}}}{\lambda }}}{{\bf{a}}_R}\left( {{\phi _R}\left( {\bf{q}} \right)} \right){\bf{a}}_{T,s}^H\left( {{\phi _{T,s}}\left( {\bf{q}} \right)} \right),
  \end{equation}
 where ${{\bf{a}}_R}\left( \phi \right)$ and ${{\bf{a}}_{T,s}}\left( \phi  \right)$ represent the receive array response of the AIRS and the transmit array response of the source node, respectively, which can be expressed as
\begin{equation}
 \setlength\abovedisplayskip{1pt}
 \setlength\belowdisplayskip{1pt}
\small
 {{\bf{a}}_R}\left( {{\phi _R}\left( {\bf{q}} \right)} \right) = {\left[ {1,{e^{ - j2\pi \bar d{{\bar \phi }_R}\left( {\bf{q}} \right)}}, \cdots ,{e^{ - j2\pi \left( {N - 1} \right)\bar d{{\bar \phi }_R}\left( {\bf{q}} \right)}}} \right]^T}, \label{receive array response at AIRS} 
  \end{equation}
\begin{equation}
 \setlength\abovedisplayskip{1pt}
 \setlength\belowdisplayskip{1pt}
\small 
{{\bf{a}}_{T,s}}\left( {{\phi _{T,s}}\left( {\bf{q}} \right)} \right) = {\left[ {1,{e^{ - j2\pi {{\bar d}_0}{{\bar \phi }_{T,s}}\left( {\bf{q}} \right)}}, \cdots ,{e^{ - j2\pi \left( {M - 1} \right){{\bar d}_0}{{\bar \phi }_{T,s}}\left( {\bf{q}} \right)}}} \right]^T}, \label{transmit array response at source node} 
\end{equation}
with ${\bar \phi _R}\left( {\bf{q}} \right) \buildrel \Delta \over =  \sin \left( {{\phi _R}\left( {\bf{q}} \right)} \right) $, ${{\bar \phi }_{T,s}}\left( {\bf{q}} \right) \buildrel \Delta \over =  \sin \left( {{\phi _{T,s}}\left( {\bf{q}} \right)} \right) $, $\bar d = \frac{d}{\lambda }$ and ${{\bar d}_0} = \frac{{{d_0}}}{\lambda }$. Note that the AIRS placement $\bf{q}$ not only affects the path loss ${\beta _{\bf{G}}}\left( {\bf{q}} \right)$, but also the AoD/AoA of the source-AIRS link. Similarly, denote ${{\phi _T}\left( {{\bf{q}},{\bf{w}}} \right)}$ as the AoD for the communication link from the AIRS to a location $\mathbf w\in \mathcal A$. Then the corresponding channel, denoted as ${{\bf{h}}^H}\left( {{\bf{q}},{\bf{w}}} \right) \in {{\mathbb{C}}^{1 \times N}}$, can be expressed as  
  \begin{equation}
   \setlength\abovedisplayskip{1pt}
 \setlength\belowdisplayskip{1pt}
{{\bf{h}}^H}\left( {{\bf{q}},{\bf{w}}} \right) = \sqrt {{\beta _{\bf{h}}}\left( {{\bf{q}},{\bf{w}}} \right)} {e^{ - j\frac{{2\pi {d_{\bf{h}}}}}{\lambda }}}{\bf{a}}_T^H\left( {{\phi _T}\left( {{\bf{q}},{\bf{w}}} \right)} \right),
  \end{equation}
where ${{\bf{a}}_T}\left( \phi \right)$ is the transmit (reflect) array response at the AIRS, which is given by 
\begin{equation}
    \setlength\abovedisplayskip{1pt}
 \setlength\belowdisplayskip{2pt}
\small
{{\bf{a}}_T}\left( {{\phi _T}\left( {{\bf{q}},{\bf{w}}} \right)} \right) = {\left[ {1,{e^{ - j2\pi \bar d{{\bar \phi }_T}\left( {{\bf{q}},{\bf{w}}} \right)}}, \cdots ,{e^{ - j2\pi \left( {N - 1} \right)\bar d{{\bar \phi }_T}\left( {{\bf{q}},{\bf{w}}} \right)}}} \right]^T}, \label{transmit array response at AIRS} 
\end{equation} 
with ${\bar \phi _T}\left( {{\bf{q}},{\bf{w}}} \right) \buildrel \Delta \over =  \sin \left( {{\phi _T}\left( {{\bf{q}},{\bf{w}}} \right)} \right)$. 
  
  Then the received signal at each location $\mathbf w\in \mathcal A$ is
  \begin{equation}
  \setlength\abovedisplayskip{1pt}
 \setlength\belowdisplayskip{1pt}
y\left( {{\bf{q}},{\bf{\Theta }},{\bf{w}},{\bf{v}}} \right) = {{\bf{h}}^H}\left( {{\bf{q}},{\bf{w}}} \right){\bf{\Theta G}}\left( {\bf{q}} \right){\bf{v}}\sqrt P s + n,
  \end{equation}
  where ${\bf{\Theta }} = {\rm{diag}}\left( {{e^{j{\theta _1}}}, \cdots ,{e^{j{\theta _N}}}} \right)$ is a diagonal phase-shift matrix with ${\theta _n} \in \left[ {0,2\pi } \right)$ denoting the phase shift of the $n$th reflection element; $P$ and $s$ are the transmit power and signal at the source node, respectively; $\bf{v}$ is the transmit beamforming vector at the source node with $\left\| {\bf{v}} \right\| = 1$; $ n \in {\rm{{\cal C}{\cal N}}}\left( {0,{\sigma ^2}} \right)$ is the additive white Gaussian noise (AWGN). The received SNR at the location $\mathbf w\in \mathcal A$ can be written as
\begin{equation}
 \setlength\abovedisplayskip{1pt}
 \setlength\belowdisplayskip{1pt}
\gamma \left( {{\bf{q}},{\bf{\Theta }},{\bf{w}},{\bf{v}}} \right) = \frac{{P{{\left| {{{\bf{h}}^H}\left( {{\bf{q}},{\bf{w}}} \right){\bf{\Theta G}}\left( {\bf{q}} \right){\bf{v}}} \right|}^2}}}{{{\sigma ^2}}}.\label{SNRAtLocationw}
 \end{equation} 
By denoting ${\bm{\theta }} = \left[ {{\theta _1}, \cdots ,{\theta _N}} \right]$, our objective is to maximize the minimum SNR within the rectangular area $\mathcal A$ (since in practice the destination nodes can be randomly located in it), by jointly optimizing the AIRS placement $\bf{q}$, the phase shifts $\bm{\theta }$ and the transmit beamforming vector $\bf{v}$. This optimization problem can be formulated as 
\begin{equation}\label{original problem}
 \setlength\abovedisplayskip{1pt}
 \setlength\belowdisplayskip{1pt}
 \begin{aligned}
\left( \rm{{P1}} \right){\rm{    }}\mathop {\max }\limits_{{\bf{q}},{\bm{\theta }},{\bf{v}}} &\  \  \mathop {\min \ \ }\limits_{{\bf{w}} \in {\rm{{\cal A}}}} \gamma \left( {{\bf{q}},{\bf{\Theta }},{\bf{w}},{\bf{v}}} \right)\\
{\rm{    s.t.       }}{\rm{  }}&\ \  {\rm{                }}0 \le {\theta _n} \le 2\pi ,\ {\rm{    }}n = 1, \cdots ,N, \\
&\ \left\| {\bf{v}} \right\| = 1. \nonumber
\end{aligned}
\end{equation}
Problem (P1) is difficult to solve optimally in general due to the following reasons. First, the objective function is the minimum SNR over a 2D area, which is difficult to express in terms of the optimization variables. Second, the optimization problem is highly non-convex and the optimization variables $\bf{q}$, $\bm{\theta}$ and $\bf{v}$ are intricately coupled with each other in the objective function. In the following, we first rigorously show that the optimal transmit beamforming vector $\mathbf v$ is simply the maximum ratio transmission (MRT) towards the AIRS, regardless of the reflected link from the AIRS to the ground. Furthermore, for the optimization of the AIRS placement $\mathbf q$ and phase shifts $\boldsymbol \theta$, we first consider the special case with one single-location SNR maximization to gain useful insights, for which the optimal solution can be obtained in closed-form. Then for the general case for area coverage enhancement, an efficient algorithm is proposed by exploiting the similarity between phase shifts optimization for IRS and analog beamforming for the conventional phase array.

\section{Proposed Solutions}
First, by exploiting the structure of the concatenated channel ${{{\bf{\tilde h}}}^H} \triangleq {{\bf{h}}^H}\left( {{\bf{q}},{\bf{w}}} \right){\bf{\Theta G}}\left( {\bf{q}} \right)$, the optimal transmit beamforming vector at the source node can be obtained in the following proposition. 

 \emph{Proposition 1:} The optimal transmit beamforming vector $\bf{v}$ to (P1) is ${{\bf{v}}^*} = \frac{{{{\bf{a}}_{T,s}}\left( {{\phi _{T,s}}\left( {\bf{q}} \right)} \right)}}{{\left\| {{{\bf{a}}_{T,s}}\left( {{\phi _{T,s}}\left( {\bf{q}} \right)} \right)} \right\|}}$.

\begin{IEEEproof} 
For any given AIRS placement $\mathbf q$, destination node location $\mathbf w$ and phase shifts $\boldsymbol \theta$, it can be shown that the optimal beamforming vector to maximize $\gamma \left( {{\bf{q}},{\bf{\Theta }},{\bf{w}},{\bf{v}}} \right)$ in (\ref{SNRAtLocationw}), denoted as ${{\bf{v}}^*}\left( {{\bf{q}},{\bf{\Theta }},{\bf{w}}} \right)$, is the eigenvector corresponding to the largest eigenvalue of the channel matrix ${\bf{\tilde h}}{{{\bf{\tilde h}}}^H}$. Furthermore, ${\bf{\tilde h}}{{{\bf{\tilde h}}}^H}$ can be simplified as 
  \begin{equation}
  \setlength\abovedisplayskip{1pt}
 \setlength\belowdisplayskip{1pt}
  \small
  \begin{split}
 &{\bf{\tilde h}}{{{\bf{\tilde h}}}^H} = {{\bf{G}}^H}\left( {\bf{q}} \right){{\bf{\Theta }}^H}{\bf{h}}\left( {{\bf{q}},{\bf{w}}} \right){{\bf{h}}^H}\left( {{\bf{q}},{\bf{w}}} \right){\bf{\Theta G}}\left( {\bf{q}} \right)\\
 &= {\beta _{\bf{G}}}\left( {\bf{q}} \right){\left| {{\bf{a}}_R^H\left( {{\phi _R}\left( {\bf{q}} \right)} \right){{\bf{\Theta }}^H}{\bf{h}}\left( {{\bf{q}},{\bf{w}}} \right)} \right|^2}{{\bf{a}}_{T,s}}\left( {{\phi _{T,s}}\left( {\bf{q}} \right)} \right){\bf{a}}_{T,s}^H\left( {{\phi _R}\left( {\bf{q}} \right)} \right).
   \end{split}
 \end{equation}
It then follows that ${\bf{\tilde h}}{{{\bf{\tilde h}}}^H}$ is a rank-one matrix, whose eigenvector is simply $ \frac{{{{\bf{a}}_{T,s}}\left( {{\phi _{T,s}}\left( {\bf{q}} \right)} \right)}}{{\left\| {{{\bf{a}}_{T,s}}\left( {{\phi _{T,s}}\left( {\bf{q}} \right)} \right)} \right\|}}$. More importantly, this eigenvector is independent of the destination node location $\mathbf w$. Thus, it is optimal regardless of $\mathbf w$ to set transmit beamforming as  $\frac{{{{\bf{a}}_{T,s}}\left( {{\phi _{T,s}}\left( {\bf{q}} \right)} \right)}}{{\left\| {{{\bf{a}}_{T,s}}\left( {{\phi _{T,s}}\left( {\bf{q}} \right)} \right)} \right\|}}$. The proof of Proposition 1 is thus completed.
\end{IEEEproof}
By substituting ${{\bf{v}}^*}\left( {{\bf{q}},{\bf{\Theta }},{\bf{w}}} \right)$ to (\ref{SNRAtLocationw}), the resulting SNR at the destination node location $\mathbf w\in \mathcal A$ can be expressed as 
\begin{equation} 
  \setlength\abovedisplayskip{1pt}
 \setlength\belowdisplayskip{1pt}
\small
 \begin{split}
&\gamma \left( {{\bf{q}},{\bf{\Theta }},{\bf{w}}} \right) = \bar P{\left| {{{\bf{h}}^H}\left( {{\bf{q}},{\bf{w}}} \right){\bf{\Theta G}}\left( {\bf{q}} \right)\frac{{{{\bf{a}}_{T,s}}\left( {{\phi _{T,s}}\left( {\bf{q}} \right)} \right)}}{{\left\| {{{\bf{a}}_{T,s}}\left( {{\phi _{T,s}}\left( {\bf{q}} \right)} \right)} \right\|}}} \right|^2}\\
 %&= \bar PM{\left| {\sum\limits_{n = 1}^N {\sqrt {{\beta _{\bf{h}}}\left( {{\bf{q}},{\bf{w}}} \right){\beta _{\bf{G}}}\left( {\bf{q}} \right)} {e^{ - j\frac{{2\pi \left( {{d_{\bf{h}}} + {d_{\bf{G}}}} \right)}}{\lambda }}}{e^{j\left( {{\theta _n} + 2\pi \left( {n - 1} \right)\bar d\left( {{{\bar \phi }_T}\left( {{\bf{q}},{\bf{w}}} \right) - {{\bar \phi }_R}\left( {\bf{q}} \right)} \right)} \right)}}} } \right|^2}\\
 &= \frac{{\bar P\beta _0^2M{{\left| {\sum\limits_{n = 1}^N {{e^{j\left( {{\theta _n} + 2\pi \left( {n - 1} \right)\bar d\left( {{{\bar \phi }_T}\left( {{\bf{q}},{\bf{w}}} \right) - {{\bar \phi }_R}\left( {\bf{q}} \right)} \right)} \right)}}} } \right|}^2}}}{{\left( {{H^2} + {{\left\| {{\bf{q}} - {\bf{w}}} \right\|}^2}} \right)\left( {{H^2} + {{\left\| {\bf{q}} \right\|}^2}} \right)}}, \label{SNR at location w}
\end{split}
\end{equation}
where $\bar P = \frac{P}{{{\sigma ^2}}}$. As a result, problem (P1) reduces to
\begin{equation} 
  \setlength\abovedisplayskip{1pt}
 \setlength\belowdisplayskip{1pt}
\begin{aligned}
 \left( {\rm{P2}} \right)\ {\rm{    }}\mathop {\max }\limits_{{\bf{q}},{\bm{\theta }}}&\  \   \mathop {\min }\limits_{{\bf{w}} \in {\rm{{\cal A}}}} \ \gamma \left( {{\bf{q}},{\bf{\Theta }},{\bf{w}}} \right)\\
{\rm{      s.t.     }}{\rm{  }}&\ \  0 \le {\theta _n} \le 2\pi ,\ {\rm{    }}n = 1, \cdots ,N. \label{problem without v} \nonumber
\end{aligned}
\end{equation}
\vspace{-0.3cm}
\subsection{Single-Location SNR Maximization} 
In this subsection, we consider the special case of (P2) with one single destination node of known location in $\mathcal A$. Denote by ${\bf{\hat w}} $ the destination node location and $D={\left\| {\bf{\hat w}}  \right\|}$ the source-destination distance. In this case, the inner minimization of the objective function in (P2) is not needed, and problem (P2) reduces to 
\begin{equation} 
  \setlength\abovedisplayskip{1pt}
 \setlength\belowdisplayskip{1pt}
\begin{aligned}
\left( {{\rm{P}}3}\right){\rm{    }}\mathop {\max }\limits_{{{\bf{q}}},{\bm{\theta }}} &\  \  \gamma \left( {{\bf{q}},{\bf{\Theta }},{\bf{\hat w}}} \right)\\
{\rm{           }}{\rm{s.t.}}{\rm{  }}&\ \ 0 \le {\theta _n} \le 2\pi ,\ {\rm{    }}n = 1, \cdots ,N. \label{single location problem} \nonumber
\end{aligned}
\end{equation}
It is not difficult to see that at the optimal solution to (P3), the different rays reflected by the AIRS elements should be coherently added at the receiver, that is, 
\begin{equation}
  \setlength\abovedisplayskip{1pt}
 \setlength\belowdisplayskip{1pt}
\theta _n^*\left( {\bf{q}} \right)= \bar \theta  - 2\pi \left( {n - 1} \right)\bar d\left( {{{\bar \phi }_T}\left( {{\bf{q}},{\bf{\hat w}}} \right) - {{\bar \phi }_R}\left( {\bf{q}} \right)} \right),\ n = 1, \cdots ,N,  \label{single location optimal phase shift}
 \end{equation}
where $\bar \theta$ is an arbitrary phase shift that is common to all reflecting elements. As a result, the received SNR at the target location is simplified to  
\begin{equation}
  \setlength\abovedisplayskip{1pt}
 \setlength\belowdisplayskip{1pt}
{\gamma_1 \left( {{\bf{q}},{\bf{\hat w}}} \right)}  = \frac{{\bar P\beta _0^2M{N^2}}}{{\left( {{H^2} + {{\left\| {{{\bf{q}}} - {\bf{\hat w}}} \right\|}^2}} \right)\left( {{H^2} + {{\left\| {{{\bf{q}}}} \right\|}^2}} \right)}}.\label{single-location SNR}
 \end{equation}
After some manipulations, problem (P2) can be reformulated as 
\begin{equation}
\vspace{-0.2cm}
\left( {\rm{P4}} \right){\rm{    }}\mathop {\min }\limits_{{{\bf{q}}}} {\rm{  }}\left( {{H^2} + {{\left\| {{{\bf{q}}} - {\bf{\hat w}}} \right\|}^2}} \right)\left( {{H^2} + {{\left\| {{{\bf{q}}}} \right\|}^2}} \right).\nonumber
 \end{equation}

 \emph{Proposition 2:}  For the single-location SNR maximization problem (P4), the optimal horizontal placement of the AIRS is     
 \begin{equation}
   \setlength\abovedisplayskip{2pt}
 \setlength\belowdisplayskip{1pt}
  {\bf{q}}^* =   {\xi ^*}\left( \rho  \right){\bf{\hat w}} ,
  \end{equation}
  with
   \begin{equation}
    \setlength\abovedisplayskip{1pt}
 \setlength\belowdisplayskip{1pt}
   \small
 {\xi ^*}\left( \rho  \right) = \left\{ \begin{aligned}
&\frac{1}{2}, \ \ \ \ \ \ \ \ \ \ \ \ \ \ \ \  \ \ \ \ \ \ \ \ \ \   \ \ \ \ \ \ \ \ \  \  \ \ {\rm{                                                     if  }}\ \rho  \ge \frac{1}{2}\\
&\frac{1}{2} - \sqrt {\frac{1}{4} - {\rho ^2}}\  {\rm{  or  }}\ \frac{1}{2} + \sqrt {\frac{1}{4} - {\rho ^2}}, \ {\rm{        otherwise}}.
\end{aligned} \right. \label{optimal placement for single-location}
  \end{equation}
 where $\xi$ is called the ratio coefficient and $\rho  = \frac{H}D$.
 
\begin{figure}[!h]
     \vspace{-0.5cm}
  \setlength{\abovecaptionskip}{-0.05cm}
  \setlength{\belowcaptionskip}{-0.3cm}
  \centering
  \centerline{\includegraphics[width=2.0in,height=1.6in]{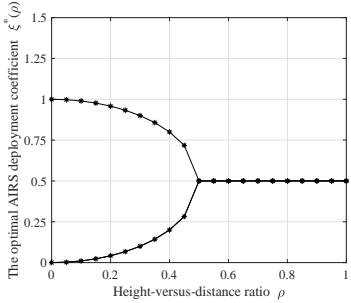}}
  \caption{The optimal AIRS deployment coefficient $  {\xi ^*}\left( \rho  \right)$ against height-versus-distance ratio $\rho$.}
  \label{coefficientVersusRatioRho}
  \vspace{-0.3cm}
  \end{figure}

\begin{IEEEproof} 
 Please refer to Appendix A.
 \end{IEEEproof} 
  Proposition 2 shows that the optimal horizontal placement of AIRS only depends on $\rho  = \frac{H}D$, that is, the ratio of AIRS height $H$ and source-destination distance $D$. For $\rho  \ge \frac{1}{2}$, the AIRS should always be placed above the middle point between the source and destination nodes. On the other hand, for $\rho<\frac{1}{2}$, there exist two optimal horizontal placement locations for the AIRS that are symmetrical about the midpoint, as shown in Fig.~\ref{coefficientVersusRatioRho}. Note that the above result is different from the conventional relay placement \cite{Relay placement for amplify-and-forward relay channels with correlated shadowing}, whose optimal solution also depends on the transmit SNR due to the relay receiver noise. 

\subsection{Area Coverage Enhancement }
Next, we study the general case of (P2) for area coverage enhancement. However, solving problem (P2) by the standard optimization techniques is difficult in general. On one hand, the AIRS placement $\mathbf q$ not only affects the link distances to/from the AIRS, but also its AoA and AoD as shown in \eqref{receive array response at AIRS}, \eqref{transmit array response at AIRS} and \eqref{SNR at location w}. On the other hand, the design of the phase shifts vector $\boldsymbol  \theta$ needs to balance the SNRs at different locations $\mathbf w$ in the target area. In this paper, by exploiting the fact that the phase shifts optimization for IRS resembles that for the extensively studied phase array or analog beamforming, we propose an efficient two-step suboptimal solution to (P2) by decoupling the phase shifts optimization from the AIRS placement design. To this end, it is noted that  problem (P2) can be equivalently written as
\begin{equation} 
 \setlength\abovedisplayskip{1pt}
 \setlength\belowdisplayskip{1pt}
\begin{aligned}
 \left( {\rm{P5}} \right)\ {\rm{    }}\mathop {\max }\limits_{{\bf{q}},{\bm{\theta }}}&\  \   \mathop {\min }\limits_{{\bf{w}} \in {\rm{{\cal A}}}} \ \frac{{{f_1}\left( {{\bf{q}},{\bm{\theta }},{\bf{w}}} \right)}}{{{f_2}\left( {{\bf{q}},{\bf{w}}} \right)}}\\
{\rm{      s.t.     }}{\rm{  }}&\ \  0 \le {\theta _n} \le 2\pi ,\ {\rm{    }}n = 1, \cdots ,N, \label{problem without v}\nonumber
\end{aligned}
\end{equation}
where ${f_1}\left( {{\bf{q}},{\bm{\theta }},{\bf{w}}} \right) \buildrel \Delta \over = {\left| {\sum\limits_{n = 1}^N {{e^{j\left( {{\theta _n} + 2\pi \left( {n - 1} \right)\bar d\left( {{{\bar \phi }_T}\left( {{\bf{q}},{\bf{w}}} \right) - {{\bar \phi }_R}\left( {\bf{q}} \right)} \right)} \right)}}} } \right|^2}$ is the {\it array gain} due to the passive beamforming by the AIRS,  and ${f_2}\left( {{\bf{q}},{\bf{w}}} \right) \buildrel \Delta \over = \left( {{H^2} + {{\left\| {{\bf{q}} - {\bf{w}}} \right\|}^2}} \right)\left( {{H^2} + {{\left\| {\bf{q}} \right\|}^2}} \right)$ accounts for the concatenated path loss.

For the proposed design, for any given AIRS placement $\mathbf q$, we design the phase shifts $\boldsymbol \theta$ in the first step to maximize the worst-case array gain by solving the following problem
\begin{equation} 
 \setlength\abovedisplayskip{1pt}
 \setlength\belowdisplayskip{1pt}
\begin{aligned}
 \left({\rm{P5.1}}\right)\   \mathop {\max }\limits_{\bm{\theta }} &\ \ \mathop {\min }\limits_{{\bf{w}} \in {\rm{{\cal A}}}}\  {f_1}\left( {{\bf{q}},{\bm{\theta }},{\bf{w}}} \right) \\
 {\rm{      s.t.     }}{\rm{  }}&\ \  0 \le {\theta _n} \le 2\pi ,\ {\rm{    }}n = 1, \cdots ,N. \label{step1 problem}\nonumber
 \end{aligned}
\end{equation}
Note that (P5.1) is an approximation of the original problem (P5) with given $\mathbf q$, since ${f_2}\left( {{\bf{q}},{\bf{w}}} \right)$ is ignored in the inner minimization of the objective function. Such an approximation is reasonable since in general, the array gain ${f_1}\left( {{\bf{q}},\bm{\theta},{\bf{w}}} \right)$ is more sensitive than the concatenated path loss ${f_2}\left( {{\bf{q}},{\bf{w}}} \right)$ to the location variation of $\mathbf w$ in the target area $\mathcal A$, especially when the source-destination distance $D \gg {D_x}$ and ${D_y}$. Then in the second step, the obtained solution to (P5.1), denoted as $ {{\bm{\theta }}^*}\left( {\bf{q}} \right)$, is substituted into the objective function of (P5). Note that even after obtaining $ {{\bm{\theta }}^*}\left( {\bf{q}} \right)$, the worst-case SNR in this target area $\mathcal A$ is still unknown, thus the AIRS placement needs to be optimized to maximize the worst-case SNR in $\mathcal A$, which can be expressed as
\begin{equation} 
 \setlength\abovedisplayskip{1pt}
 \setlength\belowdisplayskip{1pt}
 \left({\rm{P5.2}}\right)\ \mathop {\max }\limits_{\bf{q}} \ \mathop {\min }\limits_{{\bf{w}} \in {\rm{{\cal A}}}} \ \frac{{{f_1}\left( {{\bf{q}},{{\bm{\theta }}^*}\left( {\bf{q}} \right),{\bf{w}}} \right)}}{{{f_2}\left( {{\bf{q}},{\bf{w}}} \right)}}.\nonumber
 \end{equation}

\subsubsection{Phase Shifts Optimization}
In order to solve (P5.1),  we first give the following result. 

 \emph{Proposition 3:}  For any AIRS sub-array with $\bar N  \leq N$ elements and placement $\bf{q}$, assuming that its phase shifts are designed to maximize the receive SNR at a  location ${{\bf{\bar w}}}$ in $\mathcal A$, then its array gain at any other location $\mathbf w$ in $\mathcal A$ is 
  \begin{equation}
   \setlength\abovedisplayskip{1pt}
 \setlength\belowdisplayskip{1pt}
{g}\left( {\Delta \bar \phi } \right) = {\left| {\frac{{\sin \left( {\pi \bar N\bar d\Delta \bar \phi } \right)}}{{\sin \left( {\pi \bar d\Delta \bar \phi } \right)}}} \right|^2}, \label{array gain}
  \end{equation}
where $\Delta \bar \phi  = {{\bar \phi }_T}\left( {{\bf{q}},{\bf{w}}} \right) - {{\bar \phi }_T}\left( {{\bf{q}},{\bf{\bar w}}} \right)$ is the deviation of the sin-AoD (also called spatial frequency) from $\bar {\bf{w}}$.

\begin{IEEEproof} 
 According to \eqref{single location optimal phase shift}, to make all reflected signals coherently combined at the location ${{\bf{\bar w}}}$, the phase shifts of the $\bar N$ elements of the AIRS are given by 
 \begin{equation}
  \setlength\abovedisplayskip{1pt}
 \setlength\belowdisplayskip{1pt}
\theta _n\left( {\bf{q}} \right) = \bar \theta  - 2\pi \left( {n - 1} \right)\bar d\left( {{{\bar \phi }_T}\left( {{\bf{q}},{\bf{\bar w}}} \right) - {{\bar \phi }_R}\left( {\bf{q}} \right)} \right),\ n = 1, \cdots , \bar N. \label{phase shift target to bar w}
\end{equation}
By substituting $\theta _n\left( {\bf{q}} \right)$ into ${f_1}\left( {{\bf{q}},{\bm{\theta }},{\bf{w}}} \right)$ with $\bar N\leq N$, we have
 \begin{equation}
  \setlength\abovedisplayskip{1pt}
 \setlength\belowdisplayskip{1pt}
{g}\left( {{\bf{q}},{\bf{w}}} \right)= {\left| {\sum\limits_{n = 1}^{\bar N} {{e^{j\left( {2\pi \left( {n - 1} \right)\bar d\left( {{{\bar \phi }_T}\left( {{\bf{q}},{\bf{w}}} \right) - {{\bar \phi }_T}\left( {{\bf{q}},{\bf{\bar w}}} \right)} \right)} \right)}}} } \right|^2} \nonumber
\end{equation}
 \begin{equation}
   \vspace{-0.2cm}
   \setlength\abovedisplayskip{1pt}
 \setlength\belowdisplayskip{1pt}
 \hspace{-1cm}
= {\left| {\frac{{\sin \left( {\pi \bar N\bar d\left( {{{\bar \phi }_T}\left( {{\bf{q}},{\bf{w}}} \right) - {{\bar \phi }_T}\left( {{\bf{q}},{{\bf{\bar{w}}}}} \right)} \right)} \right)}}{{\sin \left( {\pi \bar d\left( {{{\bar \phi }_T}\left( {{\bf{q}},{\bf{w}}} \right) - {{\bar \phi }_T}\left( {{\bf{q}},{{\bf{\bar{w}}}}} \right)} \right)} \right)}}} \right|^2}.
\end{equation}
By letting $\Delta \bar \phi  = {{\bar \phi }_T}\left( {{\bf{q}},{\bf{w}}} \right) - {{\bar \phi }_T}\left( {{\bf{q}},{\bf{\bar w}}} \right)$, the proof of Proposition 3 is thus completed.
 \end{IEEEproof} 
    \begin{figure}[!t]
  \setlength{\abovecaptionskip}{-0.3cm}
  \setlength{\belowcaptionskip}{-0.3cm}
  \centering
  \centerline{\includegraphics[width=3.0in, height=2.0in]{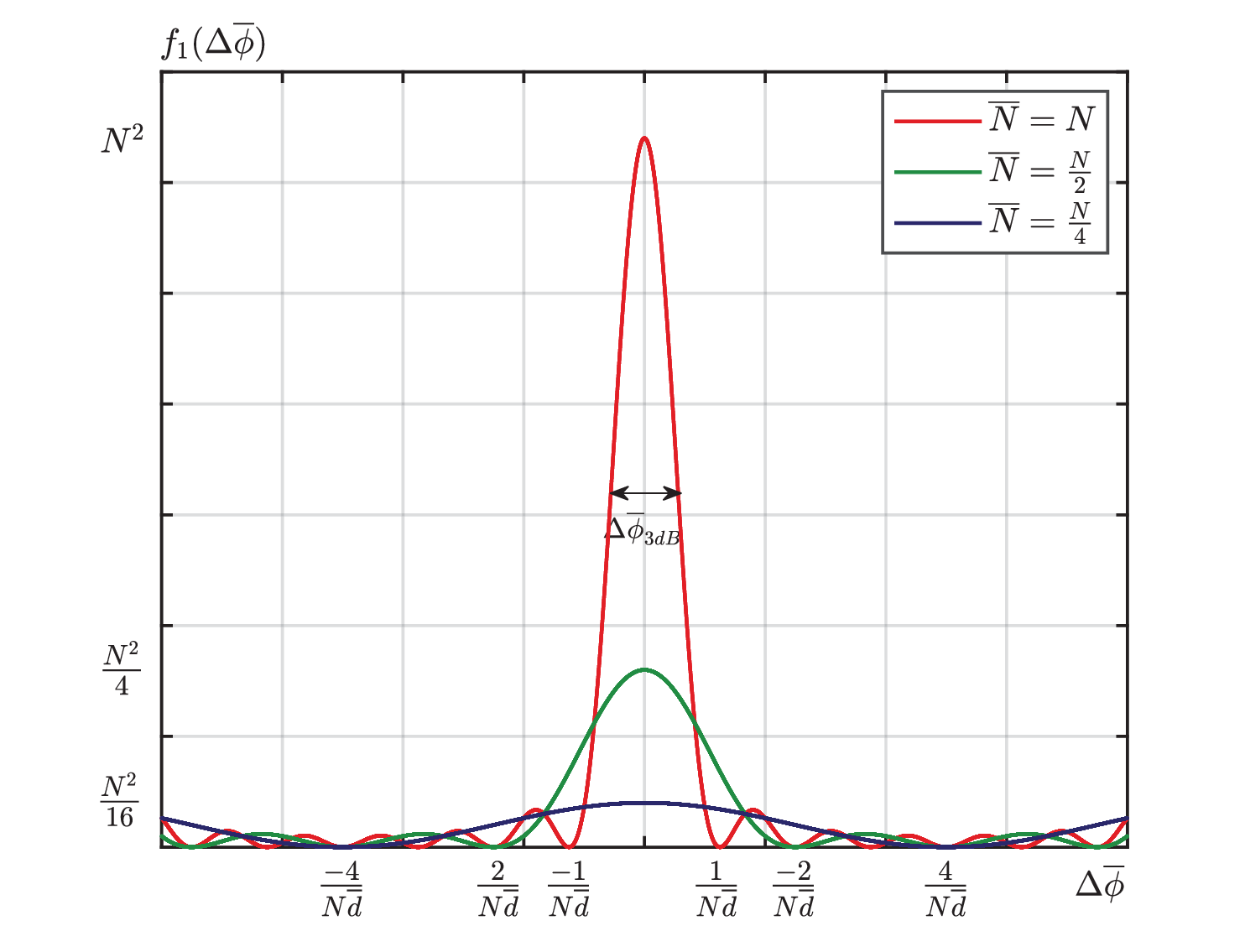}}
  \caption{AIRS array gain versus spatial frequency deviation $\Delta \bar \phi $.}
  \label{power distribution in different directions}
  \vspace{-0.5cm}
  \end{figure}
 Fig.~\ref{power distribution in different directions} shows the array gain in \eqref{array gain} versus the deviation $\Delta \bar \phi$ for different $\bar N$. It is observed that at the target location ${\bf{\bar w}}$, the received power is magnified by ${{\bar N}^2}$ times, which is in accordance with the single-location SNR maximization in (\ref{single-location SNR}). As $\Delta \bar {\phi}$ increases, the power reduces in general. By setting $\pi \bar N\bar d\Delta \bar{\phi}  = k\pi ,k = 1, \cdots ,\bar N - 1$, we have the array gain nulls, and the beamwidth can be obtained by letting $k=1$, i.e., $\Delta {\bar{\phi} _{BW}}= \frac{2}{{\bar N\bar d}}$. This reflects the well-known fact that the beamwidth of phase array is inversely proportional to the array aperture $\bar N\bar d$. Furthermore, the half-power beamwidth is known as 3-dB beamwidth, i.e., the deviation $\Delta \bar{\phi}$ at which the array gain ${g}\left( {\Delta \bar \phi } \right)$ drops to half of its peak value. According to \cite{Phased array antennas}, for a large $\bar N$, the 3-dB beamwidth can be approximated as $\Delta {\bar{\phi} _{3{\rm{dB}}}} \approx \frac{{0.8858}}{{\bar N\bar d}}$.

For the given AIRS with a total of $N$ elements and placement $\mathbf q$, assume that the phase shifts $\boldsymbol \theta$ of all the $N$ elements are designed to maximize the SNR at ${\bf{\bar w}} = {{\bf{w}}_0}$, i.e., the center of the rectangular area $\mathcal A$. Then the maximum spatial frequency deviation $\Delta \bar{\phi}$ in $\mathcal A$ can be derived as 
\begin{equation} 
 \setlength\abovedisplayskip{1pt}
 \setlength\belowdisplayskip{1pt}
\Delta {\bar{\phi} _{\max }}(\mathbf q) = \mathop {\max }\limits_{{\bf{w}} \in {\rm{{\cal A}}}} \ \left( {\left| {{{\bar \phi }_T}\left( {{\bf{q}},{\bf{w}}} \right) - {{\bar \phi }_T}\left( {{\bf{q}},{{\bf{w}}_0}} \right)} \right|} \right).
\end{equation} 

Intuitively, to achieve SNR enhancement for the entire area $\mathcal A$, the 3-dB beamwidth of the AIRS should be sufficiently large so that all locations in $\mathcal A$ lie within the main lobe of the AIRS, i.e., $\frac{{\Delta {{\bar \phi }_{3{\rm{dB}}}}}}{2}\ge \Delta {\bar{\phi} _{\max }}(\mathbf q)$. Particularly, it is observed from Fig.~\ref{power distribution in different directions}  that the 3-dB beamwidth can be increased by reducing $\bar N$ of the sub-array. This, however, decreases the peak gain of the sub-array. Therefore, there exists a design trade-off for the partition of $N$ reflecting elements into sub-arrays. To this end, we need to consider two cases depending on whether $\frac{{\Delta {{\bar \phi }_{3{\rm{dB}}}}}}{2} \ge \Delta {\bar{\phi} _{\max }}(\mathbf q)$ holds, to determine whether the sub-array architecture should be used.

\emph{Case 1:} When $\frac{{\Delta {{\bar \phi }_{3{\rm{dB}}}}}}{2} \ge \Delta {\bar{\phi} _{\max }}(\mathbf q)$, the 3-dB beamwidth can cover the entire area. Therefore, the AIRS with the full array architecture should be used for maximal area coverage. In this case, the optimal phase shifts ${{\bm{\theta }}^*}\left( {\bf{q}} \right)$ for (P5.1) are obtained by setting ${\bf{\bar w}} = {{\bf{w}}_0}$ in \eqref{phase shift target to bar w}.

\begin{figure}[!t]
    \setlength{\abovecaptionskip}{-0.05cm}
  \setlength{\belowcaptionskip}{-0.3cm}
  \centering
  \subfigure[Spatial frequency]{
    \includegraphics[width=2.3in, height=1.4in]{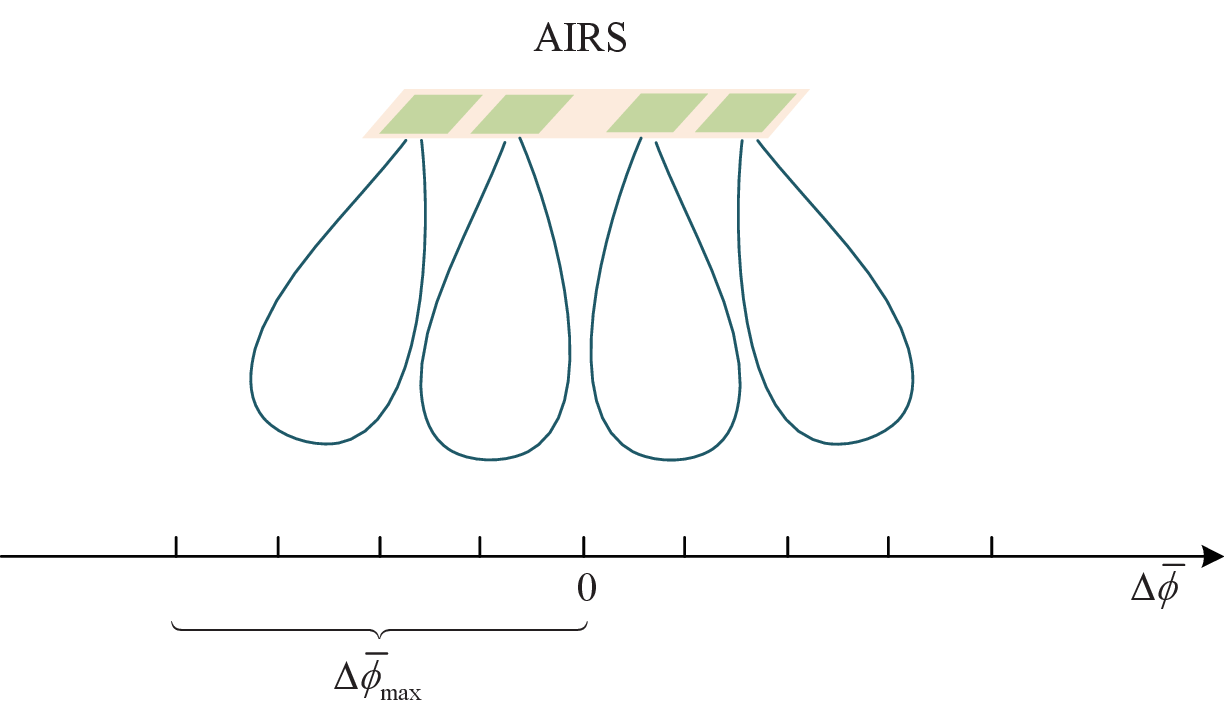}
  }
  \hspace{10mm}
  \subfigure[Horizontal location]{
    \includegraphics[width=2.3in, height=1.2in]{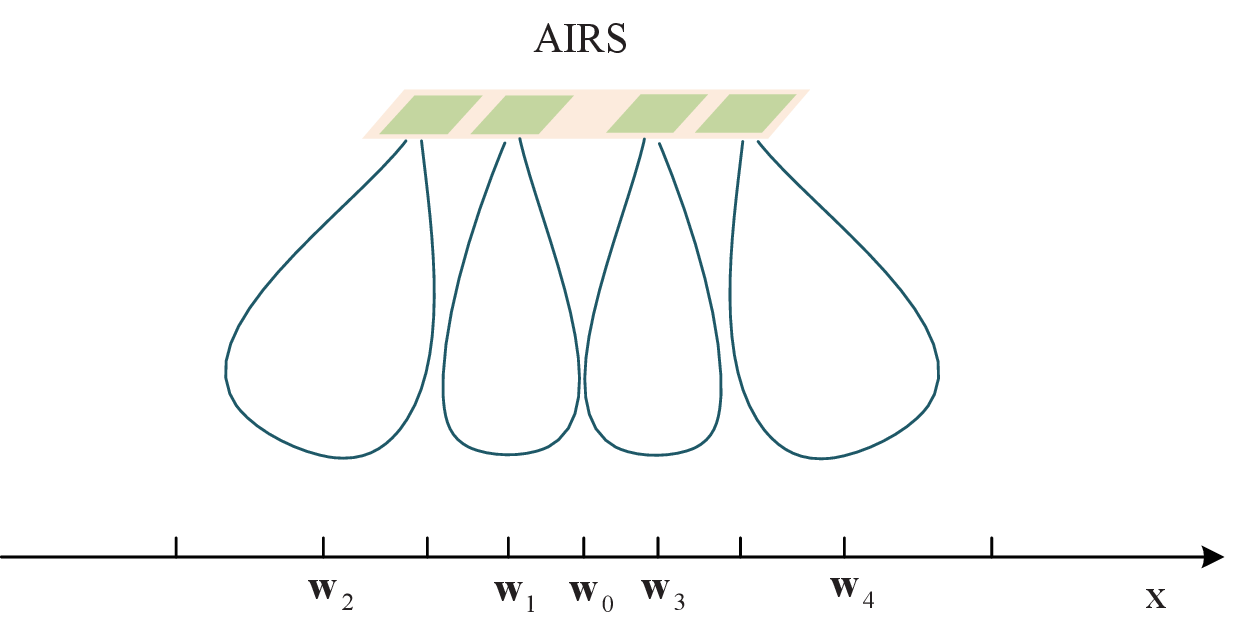}
  }
  \caption{An illustration of sub-array partition with respect to spatial frequency deviation and horizontal location, respectively.}
  \label{An illustration of sub-array partition with respect to spatial frequency deviation and physical location}
    \vspace{-0.3cm}
  \end{figure}

\emph{Case 2:} When $\frac{{\Delta {{\bar \phi }_{3{\rm{dB}}}}}}{2} < \Delta {\bar{\phi} _{\max }}(\mathbf q)$, the resulting 3-dB beamwidth using the full array architecture cannot cover the entire area. To tackle this problem, a sub-array architecture of the AIRS is proposed in this case. Specifically, the maximum spatial frequency deviation $\Delta {\bar{\phi} _{\max }}(\mathbf q)$ and the full array with $N$ elements are both equally partitioned into $L$ parts resulting in $L$ sub-arrays, each to serve one sub-area corresponding to one of the $L$ spatial frequency partitions, as illustrated in Fig.~\ref{An illustration of sub-array partition with respect to spatial frequency deviation and physical location}. The equal maximum spatial frequency deviation for each partition, denoted as $\Delta {\bar{\phi} _{\max ,l}}\left( {\bf{q}} \right), l = 1, \cdots ,L$, is reduced by $L$ times, whereas the 3-dB beamwidth of each sub-array with $N/L$ (assumed to be an integer) elements is increased by $L$ times. Notice that when reducing the number of elements, the peak sub-array gain is also reduced, as shown in Fig.~\ref{power distribution in different directions}. We set $L$ as the minimum integer to ensure $\frac{{\Delta {{\bar \phi }_{\max }}\left( {\bf{q}} \right)}}{L} \le \frac{{\Delta {{\bar \phi }_{3{\rm{dB}}}}}}{2}L$. Since $D_x\geq D_y$, by adjusting the phase shifts of each sub-array to achieve coherent signal superposition at the corresponding horizontal location in $\mathcal A$, the 3-dB beamwidth of each sub-array can cover the sub-area with spatial frequency deviation $\Delta {\bar{\phi} _{\max ,l}}\left( {\bf{q}} \right)$. The phase shifts ${{\bm{\theta }}^*}\left( {\bf{q}} \right)$ for problem (P5.1) are then obtained according to these locations.

\subsubsection{AIRS Placement Optimization}   
With the above obtained phase shifts ${{\bm{\theta }}^*}\left( {\bf{q}} \right)$ at a given AIRS placement $\bf{q}$, the worst-case SNR in $\mathcal A$ can be obtained, which occurs at the boundary point $\left( {{x_0} + \frac{{{D_x}}}{2},0} \right)$ for both the cases of full array and sub-array architecture, which has the smallest array gain but the largest concatenated path loss.  Based on the obtained worst-case SNR for any given $\bf{q}$, the AIRS placement $\bf{q}$ is then optimized. It is observed that the maximum spatial frequency deviation $\Delta {\bar{\phi} _{\max }}(\mathbf q)$ depends on $\bf{q}$, and different array architectures should be used according to the relationship between $\Delta {\bar{\phi} _{\max }}(\mathbf q)$ and $\Delta {\bar{\phi} _{3{\rm{dB}}}}$, as shown in the above. However, since it is difficult to obtain the closed-form expression of the objective function of (P5.2) for any given $\bf{q}$, (P5.2) cannot be analytically solved. Fortunately, since the optimal placement of AIRS should lie in the x-axis, i.e., ${{\bf{q}}} = {\left[ {{x},0} \right]^T}$, the horizontal placement $x$ can be found via the one-dimensional search. Thus, (P5.2) is solved and a suboptimal solution is obtained for (P5). 
    \vspace{-0.1cm}
    \begin{figure*}[htbp]
  \setlength{\abovecaptionskip}{-0cm}
  \setlength{\belowcaptionskip}{-0cm}
  \centering
\begin{minipage}[t]{0.33\linewidth}
\centering
\includegraphics[width=2.2in,height=1.76in]{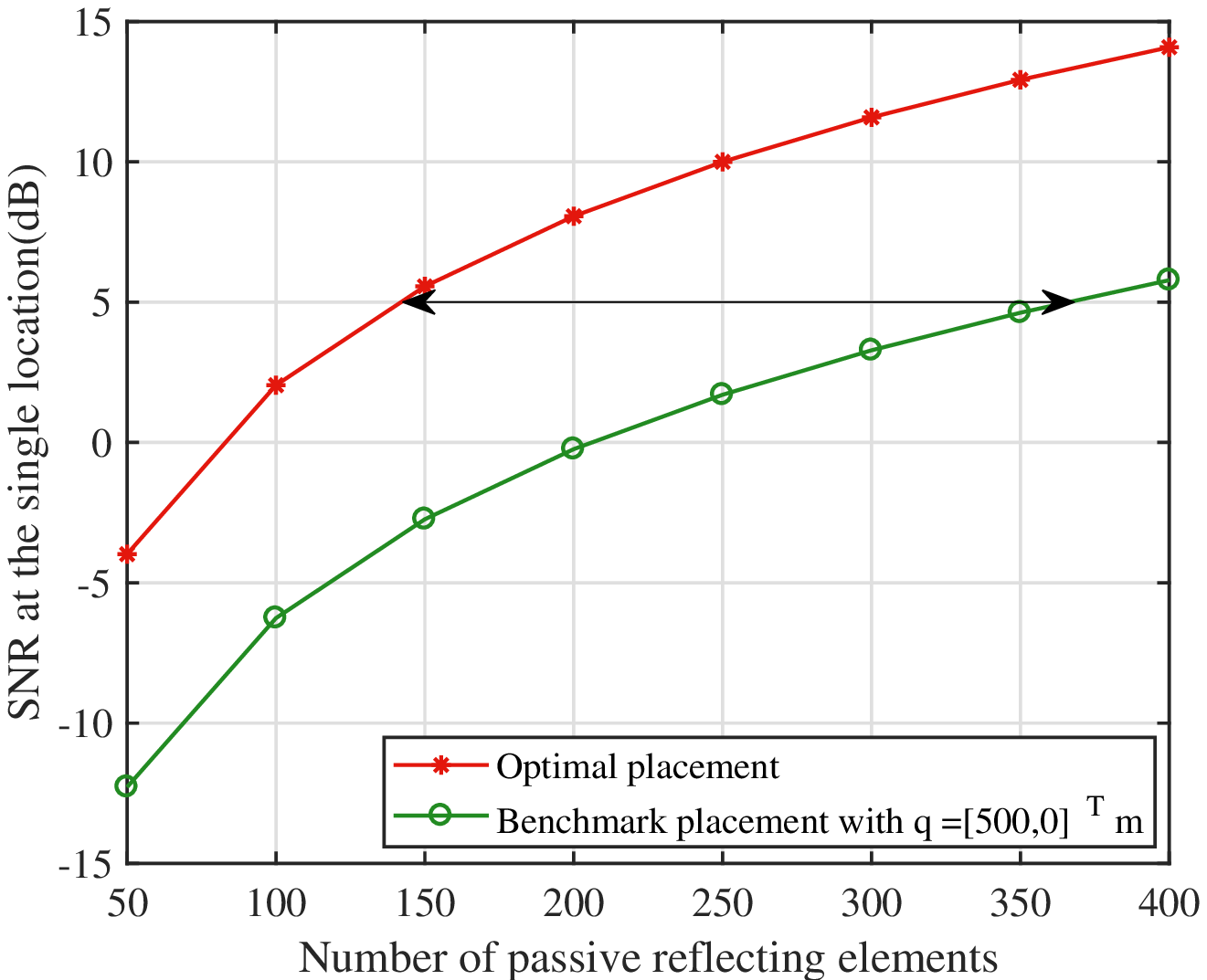}
\caption{SNR versus number of AIRS \protect\\ reflecting elements.}
  \label{SNR at the single location versus the number of passive reflecting elements}
\end{minipage}%
\begin{minipage}[t]{0.33\linewidth}
\centering
\includegraphics[width=2.2in,height=1.76in]{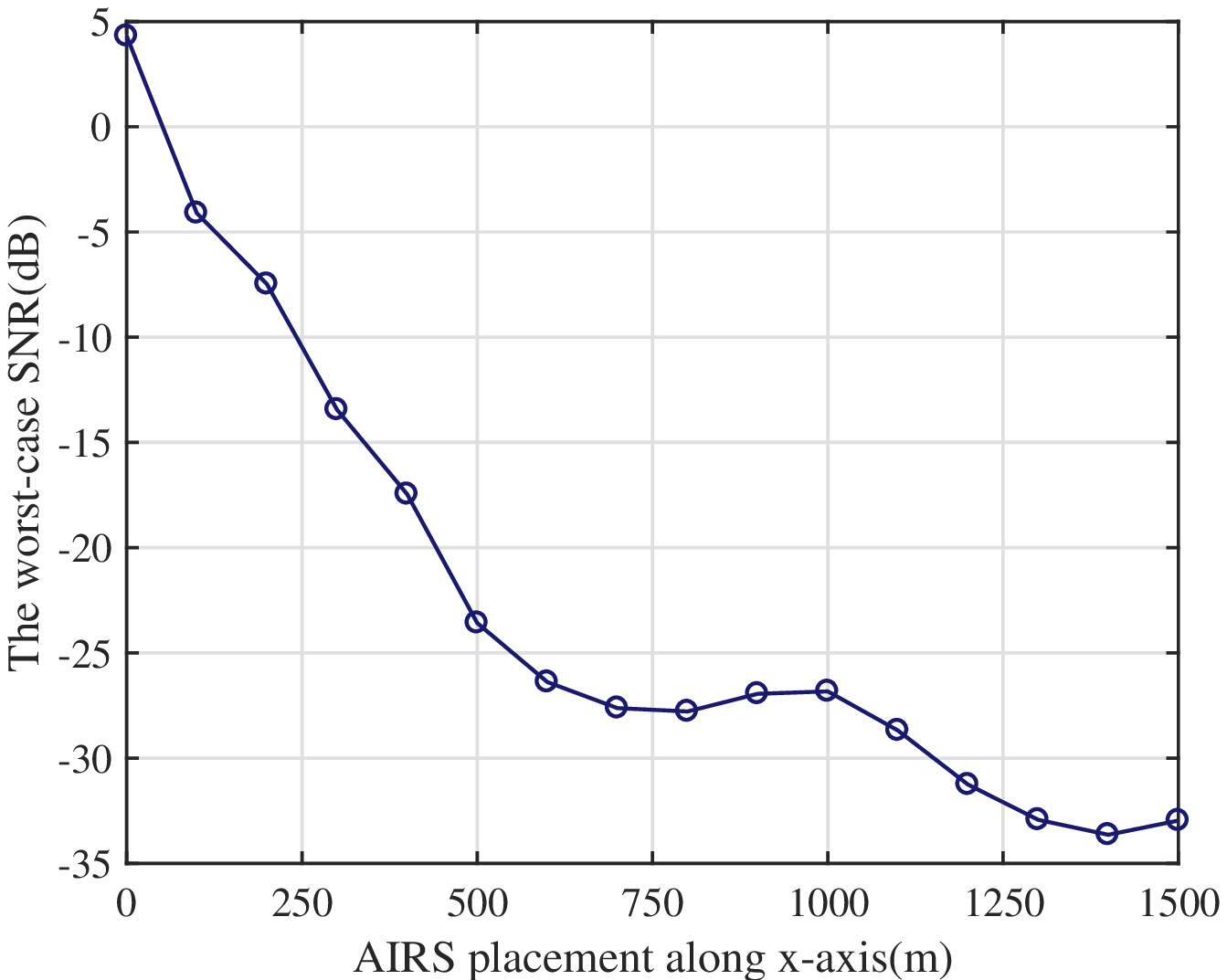}
\caption{The worst-case SNR versus AIRS \protect\\ placement along the x-axis.}
  \label{The worst-case SNR versus AIRS placement along x-axis}
\end{minipage}
\begin{minipage}[t]{0.33\linewidth}
\centering
\includegraphics[width=2.2in,height=1.76in]{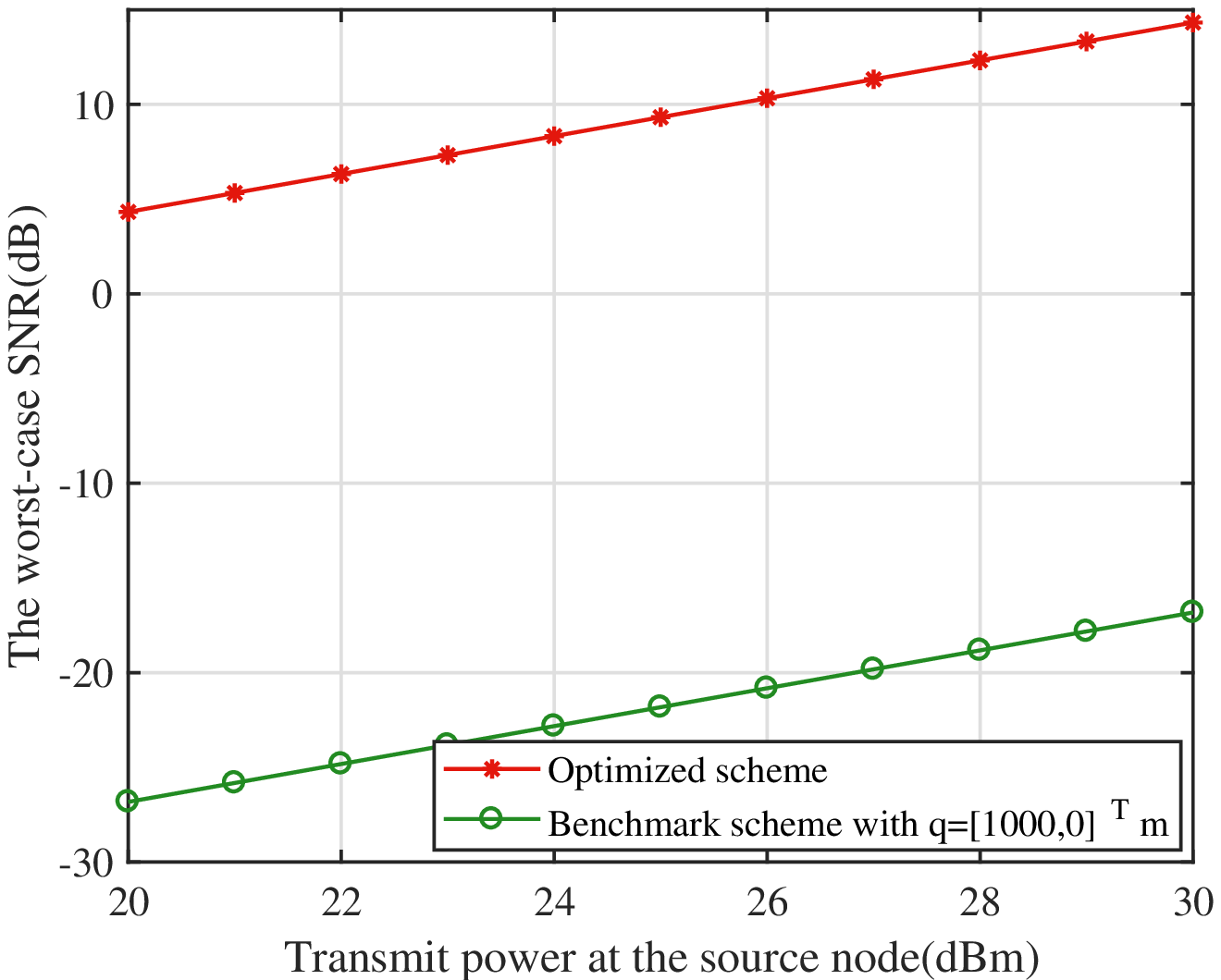}
  \caption{The worst-case SNR versus transmit \protect\\ power at the source node.}
  \label{The worst-case SNR versus transmit power at the source node}
\end{minipage}
\vspace{-0.5cm}
\end{figure*}
\section{Numerical Results}
In this section, numerical results are provided to evaluate the performance of our proposed design. The altitude of AIRS is set as $H=100$~m. The length and width of the rectangular area are $D_x=1000$~m and $D_y=600$~m, respectively, and the center is located at $(1000,0,0)$~m. Unless otherwise stated, the noise and transmit power are set as ${\sigma ^2} =  - 110$~dBm and $P=20$~dBm, respectively, and the reference channel power is ${\beta _0} =  - 40$~dB. The number of transmit antennas at source node is $M=16$. Furthermore, the separation of antennas at the source node and that of reflecting elements at the AIRS are $d_0 = \lambda /2$ and $d = \lambda /10$, respectively.

 Fig.~\ref{SNR at the single location versus the number of passive reflecting elements} shows the achievable SNR for the special case of single-location SNR maximization versus the number of passive reflecting elements. The single target location for SNR enhancement is set as $[1000,0]^T$~m. We also consider the benchmark placement with the AIRS placed above the midpoint between the source and the target location, i.e., $[500,0]^T$~m. For both the optimal placement and benchmark placement, the optimal phase shifts are applied at the AIRS to achieve coherent signal superposition at the target location. It is observed that the achievable SNR increases with the number of passive reflecting elements for both placements, as expected. In addition, the performance of the optimal placement significantly outperforms that of the benchmark placement, which shows the great benefit of placement optimization of the AIRS. For example, to achieve a target SNR of 5~dB, the number of elements required for the benchmark placement is about 360, while this number is significantly reduced to 140 for the optimal placement. 
 
For the more general area coverage or min-SNR maximization problem, Fig.~\ref{The worst-case SNR versus AIRS placement along x-axis} plots the worst-case SNR versus AIRS placement along the x-axis. The number of reflecting elements is set as $N=200$. It is observed that different from that for single-location SNR maximization case, as the AIRS moves from the source node to the target area, the performance degrades in general, although there is some small fluctuation. This can be explained by the following two reasons. First, the maximum spatial frequency deviation (or angle separation) is relatively small when the AIRS is far away from the target area, and in this case, it is more likely that the 3-dB beamwidth of the AIRS with the full array architecture is sufficient to cover the entire target area. Second, when the AIRS is near the source node, the concatenated path loss is small (see Proposition 2 and Fig. \ref{coefficientVersusRatioRho}). In contrast, when the AIRS is close to the target area, the maximum spatial frequency deviation increases significantly, thus the sub-array architecture needs to be applied to achieve area coverage by sacrificing the peak gain of each sub-array (although the concatenated path loss is similar to the case when the AIRS is near the source node). As a result, it is optimal to place the AIRS above the source node. 
  
 Last, Fig.~\ref{The worst-case SNR versus transmit power at the source node} shows the worst-case SNR versus transmit power at the source node. The number of the AIRS elements is also set as $N=200$. For comparison, we consider in this case the benchmark scheme with the AIRS placed above the center of the rectangular area, i.e., ${\bf{q}}=[1000,0]^T$~m. It is observed that the optimal AIRS placement above the source node (see Fig.~\ref{The worst-case SNR versus AIRS placement along x-axis}) significantly outperforms the benchmark placement. The above results show the importance of our proposed joint AIRS deployment and active/passive beamforming design.

\section{Conclusion}
This paper proposed a new 3D networking architecture with the AIRS to achieve efficient signal coverage from the sky. The worst-case SNR in a given target area was maximized by jointly optimizing the transmit beamforming, AIRS placement and phase shifts. We first investigated the special case of single-location SNR maximization and derived the optimal AIRS placement in closed-form, which depended on the ratio of AIRS height and source-destination distance only. Then for the general case of area coverage, we proposed an efficient suboptimal solution based on the sub-array design. Numerical results demonstrated that the proposed design can significantly improve the performance over heuristic AIRS deployment schemes.
%\begin{comment}
\begin{appendices}

\section{Proof of Proposition 2}
Denote ${{\bf{\bar q}}}$ as the projection of reference point on the straight line connecting the source node with the target location. It can be shown that ${\left\| {{{\bf{q}}} - {\bf{\hat w}}} \right\|^2} = {\left\| {{{{\bf{\bar q}}}} - {\bf{\hat w}}} \right\|^2} + {\left\| {{{\bf{q}}} - {{{\bf{\bar q}}}}} \right\|^2}$, and ${\left\| {{{\bf{q}}}} \right\|^2} = {\left\| {{{{\bf{\bar q}}}}} \right\|^2} + {\left\| {{{\bf{q}}} - {{{\bf{\bar q}}}}} \right\|^2}$. Obviously, by letting ${\left\| {{{\bf{q}}} - {{{\bf{\bar q}}}}} \right\|^2} = 0$, i.e., ${\bf{q}} = \xi {\bf{\hat w}}$, the minimum ${\left\| {{{\bf{q}}} - {\bf{\hat w}}} \right\|^2}$ and ${\left\| {{{\bf{q}}}} \right\|^2}$ can be obtained, where $\xi$ is the ratio coefficient. Thus, problem (P4) can be further reduced to:
  \begin{equation}
    \setlength\abovedisplayskip{1pt}
 \setlength\belowdisplayskip{1pt}
 \mathop {\min }\limits_\xi  \ {\left\| {{\bf{\hat w}}} \right\|^4}\left( {{\xi ^2} + {\rho ^2}} \right)\left( {{{\left( {\xi  - 1} \right)}^2} + {\rho ^2}} \right), \label{(P4) equivalent problem}
  \end{equation}
  where $\rho  = \frac{H}{{\left\| {{\bf{\hat w}}} \right\|}}$. Defining $f\left( \xi  \right) = \left( {{\xi ^2} + {\rho ^2}} \right)\left( {{{\left( {\xi  - 1} \right)}^2} + {\rho ^2}} \right)$, the first-order derivative of $f\left( \xi  \right)$ can be expressed as 
  \begin{equation}
    \setlength\abovedisplayskip{1pt}
 \setlength\belowdisplayskip{1pt}
f'\left( \xi  \right) = 4{\xi ^3} - 6{\xi ^2} + \left( {2 + 4{\rho ^2}} \right)\xi  - 2{\rho ^2}. \label{first-order derivative}
   \end{equation}
   By substituting $\xi  = \zeta  + \frac{1}{2}$ to \eqref{first-order derivative}, we have
   \begin{equation}
     \setlength\abovedisplayskip{1pt}
 \setlength\belowdisplayskip{1pt}
   f'\left( \zeta  \right) = {\zeta ^3} + a\zeta  + b,
   \end{equation}
   where $a = {\rho ^2} - \frac{1}{4}$ and $b = 0$. 
 According to the value of $\Delta  = {\left( {\frac{b}{2}} \right)^2} + {\left( {\frac{a}{3}} \right)^3} = {\left( {\frac{{{\rho ^2}}}{3} - \frac{1}{{12}}} \right)^3}$, known as the discriminant of the cubic equation, the solutions to $f'\left( \zeta  \right) = {\zeta ^3} + a\zeta  + b = 0$ can be obtained under the following three cases.
  
\begin{figure}[!h]
  \centering
  \centerline{\includegraphics[width=2.0in,height=0.8in]{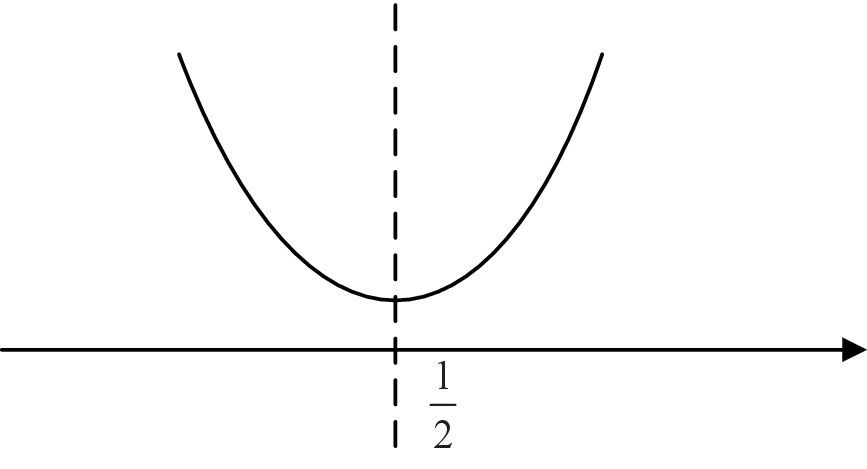}}
  \caption{The monotonicity of $f\left( \xi  \right) $ in Case 1.}
  \label{monotonicity in case 1}
  \end{figure}

  \emph{Case 1:}   When $\Delta  > 0$, that is, $\rho  > \frac{1}{2}$, there is only one real solution, which is given by

  \begin{equation}
    \setlength\abovedisplayskip{1pt}
 \setlength\belowdisplayskip{1pt}
  \small
 \zeta  = \sqrt[3]{{ - \frac{b}{2} + \sqrt {{{\left( {\frac{b}{2}} \right)}^2} + {{\left( {\frac{a}{3}} \right)}^3}} }} + \sqrt[3]{{ - \frac{b}{2} - \sqrt {{{\left( {\frac{b}{2}} \right)}^2} + {{\left( {\frac{a}{3}} \right)}^3}} }} = 0.
  \end{equation}
 then $\xi  = \zeta + \frac{1}{2}= \frac{1}{2}$. Furthermore, by checking the second-order derivative of $f\left( \xi  \right)$, we have
  \begin{equation}
    \setlength\abovedisplayskip{1pt}
 \setlength\belowdisplayskip{1pt}
   \small
   \begin{split}
f''\left( \xi  \right) &= 12\left( {{\xi ^2} - \xi  + \frac{1}{6} + \frac{{{\rho ^2}}}{3}} \right) > 12\left( {{\xi ^2} - \xi  + \frac{1}{6} + \frac{1}{{12}}} \right) \\
&= 12{\left( {\xi  - \frac{1}{2}} \right)^2} \ge 0.
  \end{split}
\end{equation}
Since the first-order derivative $f'\left( \xi  \right) $ a monotonically increasing function of $\xi$ and $f'\left( {\frac{1}{2}} \right) = 0$, the monotonicity of $f\left( \xi  \right) $ in this case is plotted  in Fig.~\ref{monotonicity in case 1}.

Therefore, $f\left( \xi  \right)$ first decreases and then increases with respect to $\xi$, and the minimum value is obtained when $\xi  = \frac{1}{2}$.

  \emph{Case 2:} When $\Delta  = 0$, namely, $\rho  = \frac{1}{2}$,  there are three equal real solutions,  i.e., ${\zeta _1} = {\zeta _2} = {\zeta _3} = 0$. Similarly, the minimum value can be obtained when $\xi  = \frac{1}{2}$.
  
  \emph{Case 3:}  When $\Delta  < 0$, there are three real solutions, which are given by 
   \begin{equation}
     \setlength\abovedisplayskip{1pt}
 \setlength\belowdisplayskip{1pt}
   \small
   \begin{split}
&{\zeta _1} = 2\sqrt { - \frac{a}{3}} \cos \frac{\vartheta }{3} = \sqrt {\frac{1}{4} - {\rho ^2}}, \\
&{\zeta _2} = 2\sqrt { - \frac{a}{3}} \cos \left( {\frac{\vartheta }{3} + 120^\circ } \right) =  - \sqrt {\frac{1}{4} - {\rho ^2}}, \\
&{\zeta _3} = 2\sqrt { - \frac{a}{3}} \cos \left( {\frac{\vartheta }{3} - 120^\circ } \right) = 0,
  \end{split}
   \end{equation}
where $\vartheta  = \arccos \frac{{ - b\sqrt { - 27a} }}{{2{a^2}}} = {90^\circ }$. Thus, ${\xi _1} = \frac{1}{2} + \sqrt {\frac{1}{4} - {\rho ^2}}$, ${\xi _2} = \frac{1}{2} - \sqrt {\frac{1}{4} - {\rho ^2}}$ and ${\xi _3} = \frac{1}{2}$. First, by letting the second-order derivative of $f''\left( \xi  \right)$ be equal to 0, we obtain 
   \begin{figure}[!t]
  \centering
  \centerline{\includegraphics[width=2.0in,height=0.8in]{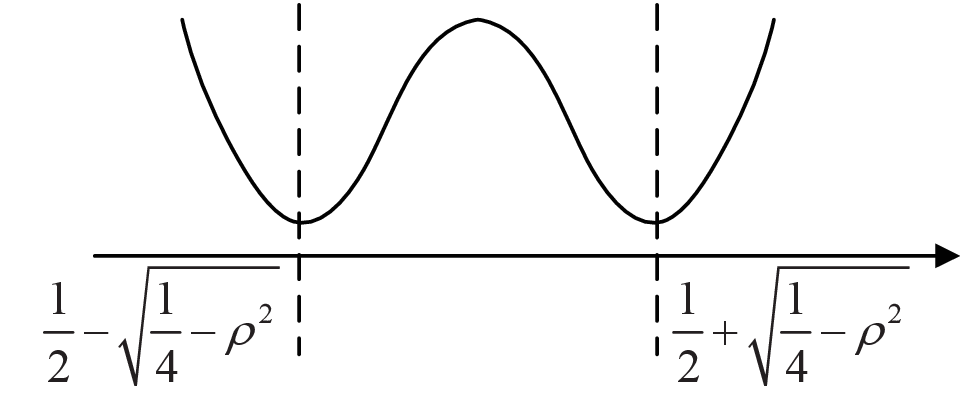}}
  \caption{The monotonicity of $f\left( \xi  \right) $ in Case 3.}
  \label{monotonicity in case 3}
  \end{figure}
 \begin{equation}
 \small
12\left( {{\xi ^2} - \xi  + \frac{1}{6} + \frac{{{\rho ^2}}}{3}} \right) = 12\left( {{{\left( {\xi  - \frac{1}{2}} \right)}^2} - \frac{1}{{12}} + \frac{{{\rho ^2}}}{3}} \right) = 0,
   \end{equation} 
Since {{\small$\xi _1^{''} = \frac{1}{2} + \sqrt {\frac{1}{{12}} - \frac{{{\rho ^2}}}{3}} $ and $\xi _2^{''} = \frac{1}{2} - \sqrt {\frac{1}{{12}} - \frac{{{\rho ^2}}}{3}}$}}, $f'\left( \xi  \right) $ increases in the interval {{\small$\left( { - \infty ,\frac{1}{2} - \sqrt {\frac{1}{{12}} - \frac{{{\rho ^2}}}{3}} } \right]$}} and {{\small$\left[ {\frac{1}{2} + \sqrt {\frac{1}{{12}} - \frac{{{\rho ^2}}}{3}} , + \infty } \right)$}}, and decreases in the interval {{\small$\left( {\frac{1}{2} - \sqrt {\frac{1}{{12}} - \frac{{{\rho ^2}}}{3}} ,\frac{1}{2} + \sqrt {\frac{1}{{12}} - \frac{{{\rho ^2}}}{3}} } \right)$}}.  
 
 Thus, $f\left( \xi  \right)$ decreases in the interval  {{\small$\left( { - \infty ,\frac{1}{2} - \sqrt {\frac{1}{4} - {\rho ^2}} } \right]$}} and {{\small$\left( {\frac{1}{2},\frac{1}{2} + \sqrt {\frac{1}{4} - {\rho ^2}} } \right]$}}, and increases in the interval {{\small$\left( {\frac{1}{2} - \sqrt {\frac{1}{4} - {\rho ^2}} ,\frac{1}{2}} \right]$ and $\left( {\frac{1}{2} + \sqrt {\frac{1}{4} - {\rho ^2}} , + \infty } \right)$}}. The monotonicity of $f\left( \xi  \right) $ in this case is plotted in Fig.~\ref{monotonicity in case 3}. Furthermore, by substituting {{\small${\frac{1}{2} - \sqrt {\frac{1}{4} - {\rho ^2}} }$}} and {{\small${\frac{1}{2} + \sqrt {\frac{1}{4} - {\rho ^2}} }$}} to $f\left( \xi  \right)$, we have {{\small$f\left( {\frac{1}{2} - \sqrt {\frac{1}{4} - {\rho ^2}} } \right) = f\left( {\frac{1}{2} + \sqrt {\frac{1}{4} - {\rho ^2}} } \right)$}}. Thus, the minimum value of \eqref{(P4) equivalent problem} is obtained when {{\small$\xi  = {\frac{1}{2} - \sqrt {\frac{1}{4} - {\rho ^2}} }$}} or {{\small$\xi  =  {\frac{1}{2} + \sqrt {\frac{1}{4} - {\rho ^2}} }$}}. The proof of Proposition 2 is thus completed.  
 \end{appendices}
\end{document}